\documentclass[aps,prl,twocolumn,tightenlines,showpacs]{revtex4}

\usepackage{amssymb}
\usepackage{epsfig}
\usepackage{amsbsy}
\usepackage{amsmath}
\usepackage{graphicx}

\begin{document}


\title{Two-Photon Transport in a Waveguide Coupled to a Cavity with a Two-level System}
\author{ T. Shi$^{1}$, Shanhui Fan$^{2}$ and C. P. Sun$^{1}$}
\affiliation{$^{1}$Institute of Theoretical Physics, Chinese
Academy of Sciences, Beijing
100190, China  \\
 $^{2}$Ginzton Laboratory, Stanford University, Stanford, California 94305, USA}

\begin{abstract}
We consider a system where a waveguide is coupled to a cavity
embedded with a two-level system (TLS), and study the effects when
a two-photon quantum state is injected into the waveguide. The
wave function of two outgoing photons is exactly solved using the
Lehmann-Symanzik-Zimmermann (LSZ) reduction formalism. Our results
explicitly exhibit the photon blockade effects in the strong
atom-cavity coupling regime. The quantum statistical characters of
the outgoing photons, including the photon bunching and
anti-bunching behaviors, are also investigated in both the strong
and weak coupling regimes. These results agree with the
observations of recent experiments.
\end{abstract}

\maketitle

\begin{figure}[tbp]
\includegraphics[bb=32 299 565 685, width=7 cm, clip]{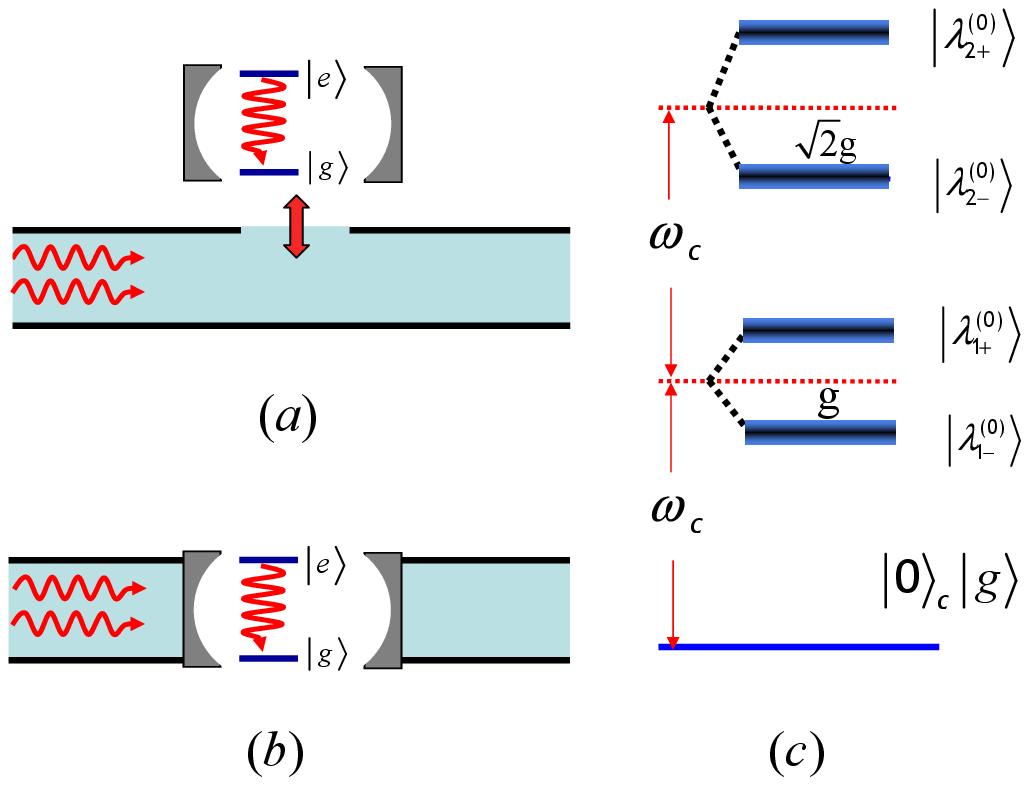}
\caption{(Color online) Two kinds of coupling structures and the schematic
of energy spectrum. (a) The side-coupled waveguide; (b) The direct-coupled
waveguide; (c) the schematic for the energy spectrum of JC model in the
sub-spaces $n=0$, $1$, and $2$.}
\label{fig1}
\end{figure}

\textit{Introduction. --- }The Jaynes-Cummings (JC) system, involving cavity
quantum electrodynamics (QED) for a two-level atom inside a cavity, is of
most importance for quantum optics \cite{QO} and its applications. In the
past decade, the JC system, in the regime where there are only a few photons
inside the cavity, has been very extensively studied due to its potential
applications for both quantum information processing and quantum device
physics \cite{KimbleNat}. The later is usually based on some solid state
systems resembling the JC system, such as the superconducting circuit QED
systems \cite{cqed} and optomechanical architectures \cite{om}. Recent
experiments about various JC systems include the observation of photon
blockade in a macroscopic cavity coupled to an atom \cite{Kimble}, the
demonstrations of on-chip cavities coupling to atom or atom-like objects
\cite{AC}, as well as the integration of such atom-cavity systems with
waveguides \cite{W-AC}. A few configurations, where the atom-cavity system
is either side-coupled \cite{fan1,zhou}, or directly coupled \cite{Kimble,B}
to a waveguide, are schematically shown in Fig. \ref{fig1}a and Fig. \ref%
{fig1}b.

In the context of these recent experiments, it is necessary to theoretically
study the intensity and coherence properties of transmitted or reflected
light, when a few-photon quantum state is injected into the system through a
waveguide. In Ref. \cite{Kimble,B}, the quantum states inside the cavity
were expanded on a basis of photon number states. Truncating the number of
basis states then reduced the Master equations to an ordinary differential
equations which can be solved numerically. This system has also been
simulated by using the quantum trajectory approach \cite{QT}. Analytically,
closed-form formulas regarding the transmission and coherence properties
have been obtained, either in the weak excitation limit where the atom is
assumed to be mostly in the ground state \cite{WE}, or in a mean-field-like
approach where the expectation value of operator product is taken as the
product of operator expectation values \cite{MF}.

In this Letter, using field-theoretic techniques \cite{Sun}, we provide an
exact analytic formula for the out-going photon wave-functions, when a
two-photon state is injected into the system shown in Fig. \ref{fig1}. Our
work is in contrast to most existing theoretical works that used coherent
state input. Given that these systems will be ultimately used to process
quantum information, understanding their response to states other than
coherent states is very important. Also, our result is exact, without the
need of either the mean-field approximation, or being restricted to the weak
excitation limit. In this system, the indirect photon-photon interaction is
strong when the atom has significant probability in the excited state \cite%
{MF}. Moreover, it is known even a single-photon pulse, when properly
designed, can completely invert an atom in this system \cite{Single}. Thus,
it is important to go beyond the weak excitation limit.

\textit{Transport model and }$S$\textit{-matrix. --- }The systems in Fig. %
\ref{fig1}a, with a waveguide side-coupled to a cavity, is described by a
Hamiltonian \cite{RS} $H=H_{\mathrm{W}}+H_{\mathrm{JC}}+H_{\mathrm{I}}$
containing three parts: (a) The waveguide Hamiltonian$\ H_{\mathrm{W}%
}=\sum_{k}\varepsilon _{k}a_{k}^{\dagger }a_{k}$, where $a_{k}$ ($%
a_{k}^{\dagger }$) denotes the annihilation (creation) operators of the
photon. $\varepsilon _{k}=v\left\vert k\right\vert $ is the waveguide
dispersion relation, and we take the speed of light $v$ as unity. (b) The JC
Hamiltonian for the coupling of the cavity field to the TLS%
\begin{equation}
H_{\mathrm{JC}}=\omega _{c}a^{\dagger }a+\Omega \left\vert e\right\rangle
\left\langle e\right\vert +g(a^{\dagger }\left\vert g\right\rangle
\left\langle e\right\vert +a\left\vert e\right\rangle \left\langle
g\right\vert ),  \label{JC}
\end{equation}%
where $a$ ($a^{\dagger }$) denotes the annihilation (creation) operators of
the photon in the cavity with frequency $\omega _{c}$, $\left\vert
e\right\rangle $ ($\left\vert g\right\rangle $) denotes the excited (ground)
state of the TLS with energy level spacing $\Omega $, and $g$ is the
coupling constant of the TLS and the cavity field. (c) The term $H_{\mathrm{I%
}}=V\sum_{k}(a_{k}^{\dagger }a+\mathrm{H.c.})/\sqrt{L}$, describing the
coupling between the cavity and the waveguide. This term results in the
decay of the cavity mode into the waveguide. Here, $V$ is the
waveguide-cavity coupling constant and $L$ is the length of waveguide.

In term of the bonding and anti-bonding waveguide mode operators defined as $%
e_{k}(o_{k})=(a_{k}\pm a_{-k})/\sqrt{2}$, we can write $H=H_{\mathrm{e}}+H_{%
\mathrm{o}}$. The bonding sub-system is described by%
\begin{equation}
H_{\mathrm{e}}=\sum_{k>0}ke_{k}^{\dagger }e_{k}+\frac{\tilde{V}}{\sqrt{L}}%
\sum_{k>0}(e_{k}^{\dagger }a+\mathrm{H.c.})+H_{\mathrm{JC}}\text{,}
\label{He}
\end{equation}%
where $\tilde{V}=\sqrt{2}V$. The anti-bonding sub-system is described by $H_{%
\mathrm{o}}=\sum_{k>0}ko_{k}^{\dagger }o_{k}$ and is decoupled from the
cavity.

Next, we utilize the Lehmann-Symanzik-Zimmermann reduction approach \cite%
{Sun}\ to study the bonding sub-system as described by the Hamiltonian $H_{%
\mathrm{e}}$. We aim to calculate the $S$-matrix elements between the
incoming and outgoing $n$-photon states, specified by their photon momenta $%
\mathbf{k}=k_{1},...,k_{n}$ and $\mathbf{p}=p_{1},...,p_{n}$, respectively.
The two-photon $S$-matrix has the form%
\begin{equation}
S_{p_{1}p_{2}k_{1}k_{2}}=S_{p_{1}k_{1}}S_{p_{2}k_{2}}+S_{p_{2}k_{1}}S_{p_{1}k_{2}}+iT_{p_{1}p_{2}k_{1}k_{2}},
\label{S2}
\end{equation}%
where $S_{pk}=\delta _{pk}+iT_{pk}$ is the single-photon $S$-matrix element.
To determine the $T$-matrix, we first exactly calculate the connected Green
function $G_{[\mathbf{p;k}]}(\omega _{\mathbf{p}},\omega _{\mathbf{k}})=\int
G_{[\mathbf{p;k}]}(\mathbf{t}^{\prime },\mathbf{t})\prod_{j=1}^{n}[-\exp
(i\omega _{p_{j}}t_{j}^{\prime }-i\omega _{k_{j}}t_{j})dt_{j}dt_{j}^{\prime
}/2\pi ]$ in the frequency domain with $\omega _{\mathbf{k}}=\omega
_{k_{1}},...,\omega _{k_{n}}$ and $\omega _{\mathbf{p}}=\omega
_{p_{1}},...,\omega _{p_{n}}$, where%
\begin{equation}
G_{[\mathbf{p;k}]}(\mathbf{t}^{\prime },\mathbf{t})=\left. \frac{\delta
^{2n}\ln Z[\eta _{k},\eta _{k}^{\ast }]}{\delta \eta _{p_{1}}^{\ast
}(t_{1}^{\prime })\text{...}\delta \eta _{p_{n}}^{\ast }(t_{n}^{\prime
})\delta \eta _{k_{1}}(t_{1})\text{...}\delta \eta _{k_{n}}(t_{n})}%
\right\vert _{\substack{ \eta _{k}=0 \\ \eta _{k}^{\ast }=0}},  \label{G}
\end{equation}%
with $\mathbf{t}=t_{1},...,t_{n}$, and $\mathbf{t}^{\prime }=t_{1}^{\prime
},...,t_{n}^{\prime }$. $Z[\eta _{k},\eta _{k}^{\ast }]=\int D[e,a,\sigma
]\exp \{i\int dt[L_{\mathrm{e}}+\sum_{k>0}(\eta _{k}^{\ast }e_{k}+h.c)]\}$
is a generating functional with $L_{\mathrm{e}}$ being the Lagrangian for
the bonding subsystem, and $\sigma $ denoting the variables of the TLS. The $%
T$-matrix elements can then be related to the Green function by%
\begin{equation}
iT_{[\mathbf{p;k}]}=\left. \frac{(2\pi )^{n}G_{[\mathbf{p;k}]}(\omega _{%
\mathbf{p}},\omega _{\mathbf{k}})}{\prod_{j=1}^{n}[G_{0}(\omega
_{p_{j}},p_{j})G_{0}(\omega _{k_{j}},k_{j})]}\right\vert _{\substack{ \omega
_{p_{j}}=p_{j} \\ \omega _{k_{j}}=k_{j}}},  \notag
\end{equation}%
where the bare Green function $G_{0}$ defined by$\ iG_{0}^{-1}(\omega
_{p},p)=\omega _{p}-\varepsilon _{p}+i0^{+}$.

For one or two-photons, this computation results in%
\begin{equation}
iT_{pk}=-\tilde{V}^{2}\int \frac{dt^{\prime }dt}{2\pi }e^{ipt^{\prime
}-ikt}\left\langle \mathcal{T}a(t^{\prime })a^{\dagger }(t)\right\rangle ,
\label{Tpk}
\end{equation}%
and%
\begin{equation}
iT_{p_{1}p_{2}k_{1}k_{2}}=\tilde{V}^{4}\int (\prod_{j=1,2}\frac{%
dt_{j}dt_{j}^{\prime }}{2\pi }e^{ip_{j}t_{j}^{\prime }-ik_{j}t_{j}})G_{4},
\label{T2}
\end{equation}%
Here, $G_{4}=\left\langle \mathcal{T}a(t_{1}^{\prime })a(t_{2}^{\prime
})a^{\dagger }(t_{1})a^{\dagger }(t_{2})\right\rangle $ is the four point
Green function of $a(t)=\exp (iH_{\mathrm{eff}}t)a\exp (-iH_{\mathrm{eff}}t)$%
, and the effective non-Hermitian Hamiltonian $H_{\mathrm{eff}}$ is obtained
by simply replacing $\omega _{c}$ with $\alpha =\omega _{c}-i\tilde{V}^{2}/2$
in $H_{\mathrm{JC}}$. The imaginary part of $\alpha $ is the cavity decay
rate. $\left\langle \mathcal{T}...\right\rangle $ is the time-ordered
average on the state $\left\vert 0\right\rangle _{\mathrm{c}}\left\vert
g\right\rangle $. Here, $\left\vert 0\right\rangle _{\mathrm{c}}$ and $%
\left\vert g\right\rangle $ denote the vacuum state of cavity field and the
ground state of TLS, respectively. Since excitation number $N=a^{\dagger
}a+\left\vert e\right\rangle \left\langle e\right\vert $ commutes with $H_{%
\mathrm{eff}}$, in its invariant subspace with $N=n$,$\ H_{\mathrm{eff}}$ is
diagonalized with the eigenstates $\left\vert \lambda _{n\pm }\right\rangle =%
\mathcal{N}_{n\pm }\{-\sqrt{n}g\left\vert n-1\right\rangle _{\mathrm{e}%
}\left\vert e\right\rangle +[\Omega +(n-1)\alpha -\lambda _{n\pm
}]\left\vert n\right\rangle _{\mathrm{e}}\left\vert g\right\rangle \}$ and
the corresponding eigenvalues $\lambda _{n\pm }=\{\Omega +(2n-1)\alpha \pm
\lbrack (\Omega -\alpha )^{2}+4ng^{2}]^{1/2}\}/2$,$\ $where $\mathcal{N}%
_{n\pm }$ are the normalization constants. The schematic for the spectrum of
JC model is shown in Fig. \ref{fig1}c. we notice that the eigenstates $%
\left\vert \lambda _{n\pm }\right\rangle $ are not orthogonal to each other.
Thus, in the following calculations, we need to use the bi-orthogonal basis
approach \cite{BO}\ with\ the eigenstates $\left\vert \lambda _{n\pm }^{\ast
}\right\rangle $ of $H_{\mathrm{eff}}^{\ast }$ corresponding to the
eigenvalues $\lambda _{n\pm }^{\ast }$. The orthogonal relations are $%
\left\langle \lambda _{n\mp }^{\ast }\left\vert \lambda _{n\pm
}\right\rangle \right. =0$ and $\left\langle \lambda _{n\pm }^{\ast
}\left\vert \lambda _{n\pm }\right\rangle \right. =1$.

In the above bi-orthogonal basis, we can evaluate the correlations $%
\left\langle \mathcal{T}a(t^{\prime })a^{\dagger }(t)\right\rangle $ and $%
G_{4}$, to obtain the single-photon and two-photon $T$-matrices and $S$%
-matrices. The results are listed as follows. \textbf{1.} For the bonding
modes, the single photon $S$-matrix is $S_{pk}=t_{k}\delta _{kp}$, where $%
t_{k}=\exp (-i2\delta _{k})$ and phase shift $\delta _{k}=\arg [(k-\lambda
_{1+})(k-\lambda _{1-})]$. The anti-bonding modes are free of\ coupling,
thus possess $S$-matrix element $S_{pk}^{(o)}=\delta _{kp}$. \textbf{2.}
Then the reflection and transmission coefficients are obtained as\ $\bar{r}%
_{k}=(t_{k}-1)/2$ and $\bar{t}_{k}=(t_{k}+1)/2$, which agree with Ref. \cite%
{RS}. \textbf{3. }The two-photon $T$-matrix elements are explicitly obtained

\begin{eqnarray}
iT_{p_{1}p_{2}k_{1}k_{2}} &=&\frac{i\tilde{V}^{4}g^{4}(E-\alpha -\Omega
)\delta _{p_{1}+p_{2},E}}{\pi \prod_{s=\pm }(E-\lambda _{2s})}  \label{TT} \\
\times  &&\frac{[(E-2\Omega )(E-2\alpha )-4g^{2}]}{\prod_{s=\pm
}\prod_{i=1,2}(k_{i}-\lambda _{1s})(p_{i}-\lambda _{1s})},  \notag
\end{eqnarray}%
where $E=k_{1}+k_{2}$ is the total energy of the incident photons. The
square norm $T_{2}=\left\vert T_{p_{1}p_{2}k_{1}k_{2}}\right\vert ^{2}$ of $T
$-matrix element exhibits the two-photon background fluorescence in the
bonding mode.
\begin{figure}[tbp]
\includegraphics[bb=25 257 579 741, width=7 cm, clip]{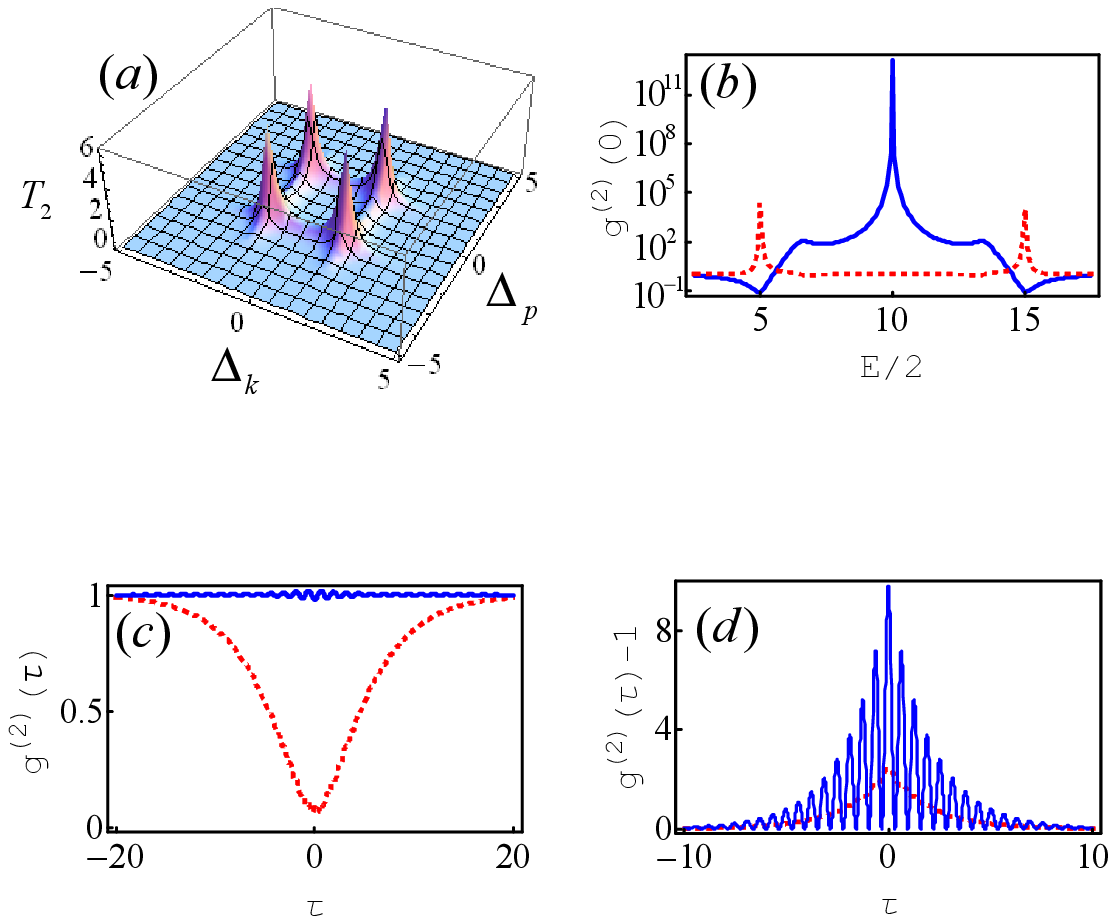}
\caption{(Color online) The two-photon background fluorescence and the
correlation functions for strong coupling regime $g=5$, the system
parameters are $\protect\omega _{c}=\Omega =10$ and $\tilde{V}$ is taken as
unit: (a) Two-photon background fluorescence for $E=\protect\lambda %
_{2s}^{(0)}$; (b) $g^{(2)}(0)$ of the reflected (solid (blue) curve) and
transmitted (dashed (red) curve) photons; (c) $g^{(2)}(\protect\tau )$ of
the reflected (dashed (red) curve) photons for $E/2=\protect\lambda %
_{1s}^{(0)}$ and transmitted (solid (blue) curve) photons for $E/2=\Omega $;
(d) $(g^{(2)}(\protect\tau )-1)/10^{15}$ of the reflected (solid (blue)
curve) photons for $E/2=\Omega $ and $(g^{(2)}(\protect\tau )-1)/10^{6}$ of
transmitted (dashed (red) curve) photons for $E/2=\protect\lambda _{1s}^{(0)}
$.}
\label{fig2}
\end{figure}

\textit{Two-photon wave functions --- }Let\textit{\ }$a_{R}(x_{1})$ ($%
a_{L}(x_{2})$) denote the annihilation operators of right (left) moving
photons \cite{Fan}. It follows from Eqs. (\ref{S2}) and (\ref{TT}) that the
out-going state $\left\vert X_{\mathrm{out}}\right\rangle =\left\vert t_{%
\mathrm{out}}\right\rangle +\left\vert r_{\mathrm{out}}\right\rangle
+\left\vert rt_{\mathrm{out}}\right\rangle $ for two incident right-moving
photons with momenta $k_{1}$ and $k_{2}$ contains three parts: (a) The
quantum state of two transmitted photons
\begin{equation*}
\left\vert t_{\mathrm{out}}\right\rangle =\int
dx_{1}dx_{2}t_{2}(x_{1},x_{2})a_{R}^{\dagger }(x_{1})a_{R}^{\dagger
}(x_{2})\left\vert 0\right\rangle \left\vert g\right\rangle
\end{equation*}%
explicitly defined by the \ two-photon wavefunction%
\begin{equation}
t_{2}(x_{1},x_{2})=\frac{1}{2\pi }e^{iEx_{c}}[\bar{t}_{k_{1}}\bar{t}%
_{k_{2}}\cos (\Delta _{k}x)-F(\lambda ,x)],  \label{t2}
\end{equation}%
where $\Delta _{k}=k_{1}-k_{2}$,%
\begin{equation}
F(\lambda ,x)=\frac{\tilde{V}^{4}g^{4}\sum_{s=\pm }s(E-2\lambda _{1s})\exp
[i(\frac{E}{2}-\lambda _{1,-s})\left\vert x\right\vert ]}{4(\lambda
_{1+}-\lambda _{1-})\prod_{s=\pm }[(E-\lambda
_{2s})\prod_{i=1,2}(k_{i}-\lambda _{1s})]},
\end{equation}%
and $x=x_{1}-x_{2}$ and $x_{c}=(x_{1}+x_{2})/2$ are the relative and center
of mass coordinates, respectively; (b) The quantum state of\ the two
reflected photons
\begin{equation*}
\left\vert r_{\mathrm{out}}\right\rangle =\int
dx_{1}dx_{2}r_{2}(x_{1},x_{2})a_{L}^{\dagger }(x_{1})a_{L}^{\dagger
}(x_{2})\left\vert 0\right\rangle \left\vert g\right\rangle
\end{equation*}%
explicitly defined by
\begin{equation}
r_{2}(x_{1},x_{2})=\frac{1}{2\pi }e^{iEx_{c}}[\bar{r}_{k_{1}}\bar{r}%
_{k_{2}}\cos (\Delta _{k}x)-F(\lambda ,x)];  \label{r2}
\end{equation}%
(c) the\ left-right entangled two-photon state
\begin{equation*}
\left\vert rt_{\mathrm{out}}\right\rangle =\int
dx_{1}dx_{2}rt_{2}(x_{1},x_{2})a_{L}^{\dagger }(x_{1})a_{R}^{\dagger
}(x_{2})\left\vert 0\right\rangle \left\vert g\right\rangle
\end{equation*}%
describes the scenario where one photon is transmitted while the other is
reflected, where%
\begin{equation}
rt_{2}=\frac{1}{2\pi }e^{i\frac{E}{2}x}[(\bar{t}_{k_{1}}\bar{r}%
_{k_{2}}e^{2i\Delta _{k}x_{c}}+\bar{t}_{k_{2}}\bar{r}_{k_{1}}e^{-2i\Delta
_{k}x_{c}})-2F(\lambda ,2x_{c})].
\end{equation}

Photon statistics can be studied through the coherence functions \cite{QO} $%
g^{(2)}(\tau )=G^{(2)}(\tau )/\left\vert G^{(1)}(0)\right\vert ^{2}$, where $%
G^{(1)}(\tau )=\left\langle F_{\mathrm{out}}\right\vert a_{F}^{\dagger
}(x+\tau )a_{F}(x)\left\vert F_{\mathrm{out}}\right\rangle $, $G^{(2)}(\tau
)=\left\langle F_{\mathrm{out}}\right\vert a_{F}^{\dagger }(x)a_{F}^{\dagger
}(x+\tau )a_{F}(x+\tau )a_{F}(x)\left\vert F_{\mathrm{out}}\right\rangle $,
and $F=R$ and $L$\ correspond to the transmitted photons and reflected
photons respectively, while $\left\vert R_{\mathrm{out}}\right\rangle
=\left\vert t_{\mathrm{out}}\right\rangle /\left\langle t_{\mathrm{out}%
}\left\vert t_{\mathrm{out}}\right\rangle \right. $ and $\left\vert L_{%
\mathrm{out}}\right\rangle =\left\vert r_{\mathrm{out}}\right\rangle
/\left\langle r_{\mathrm{out}}\left\vert r_{\mathrm{out}}\right\rangle
\right. $. We have $g^{(2)}(\tau )=C(\tau )/D$, where $C(\tau )=\left\vert
t_{2}(x,x+\tau )\right\vert ^{2}$ for photons in transmission, and$\
\left\vert r_{2}(x,x+\tau )\right\vert ^{2}$ for photons in reflection,
which is independent of $x$ and $D$ is the normalization constant.

\textit{Strong coupling regime. --- }To explore the physical consequence of
the result above, we first consider the strong coupling regime with $g>%
\tilde{V}^{2}$. The two-photon background fluorescence $T_{2}$ and the
correlation function $g^{(2)}(\tau )$ are shown in Fig. \ref{fig2} on
resonance, i.e., $\omega _{c}=\Omega $, and we assume the same energy for
the two incident photons, i.e. $\Delta _{k}=0$. When the average energy of
two photon $E/2=\lambda _{1\pm }^{(0)}$, $T_{2}$ has one\ sharp\ peak at $%
\Delta _{k}=\Delta _{p}=0$, and the four peaks emerge (see Fig. \ref{fig2}a)
for $E=\lambda _{2\pm }^{(0)}$.

In Fig. \ref{fig2}b, $g^{(2)}(0)$ as the functions of energy $E/2$ per
photon are shown for two reflected and transmitted photons, respectively.
The two reflected photons exhibit sub-Poissonian statistics ($g^{(2)}(0)<<1$%
) for $E/2=\lambda _{1\pm }^{(0)}$, and super-Poissonian statistics ($%
g^{(2)}(0)>>1$) for $E/2=\lambda _{2\pm }^{(0)}/2$ or $\Omega $. For $%
E/2=\lambda _{1\pm }^{(0)}$, the correlation functions $g^{(2)}(\tau )$ of
two reflected (dashed (red) curve in Fig. \ref{fig2}c) and transmitted
(dashed (red) curve in Fig. \ref{fig2}d) photons exhibit anti-bunching and
bunching behaviors, respectively. When $E/2=\Omega $, $g^{(2)}(\tau )$ of
two reflected (solid (blue) curve in Fig. \ref{fig2}d) and transmitted
(solid (blue) curve in Fig. \ref{fig2}c) photons exhibit large bunching and
anti-bunching behaviors, respectively. Our results for the two reflected
photons in the case of a side-coupled cavity exactly agree with that for the
transmitted photons observed in the experiment \cite{Kimble} for the case of
a directly coupled cavity as shown in Fig. \ref{fig1}b. This agreement is
not accidental,\ since the transmitted photon states in the direct-coupled
case can be mapped into the reflected photon states in the side-coupled
case, using a transformation detailed in Ref. \cite{RS}.
\begin{figure}[tbp]
\includegraphics[bb=17 290 579 741, width=7 cm, clip]{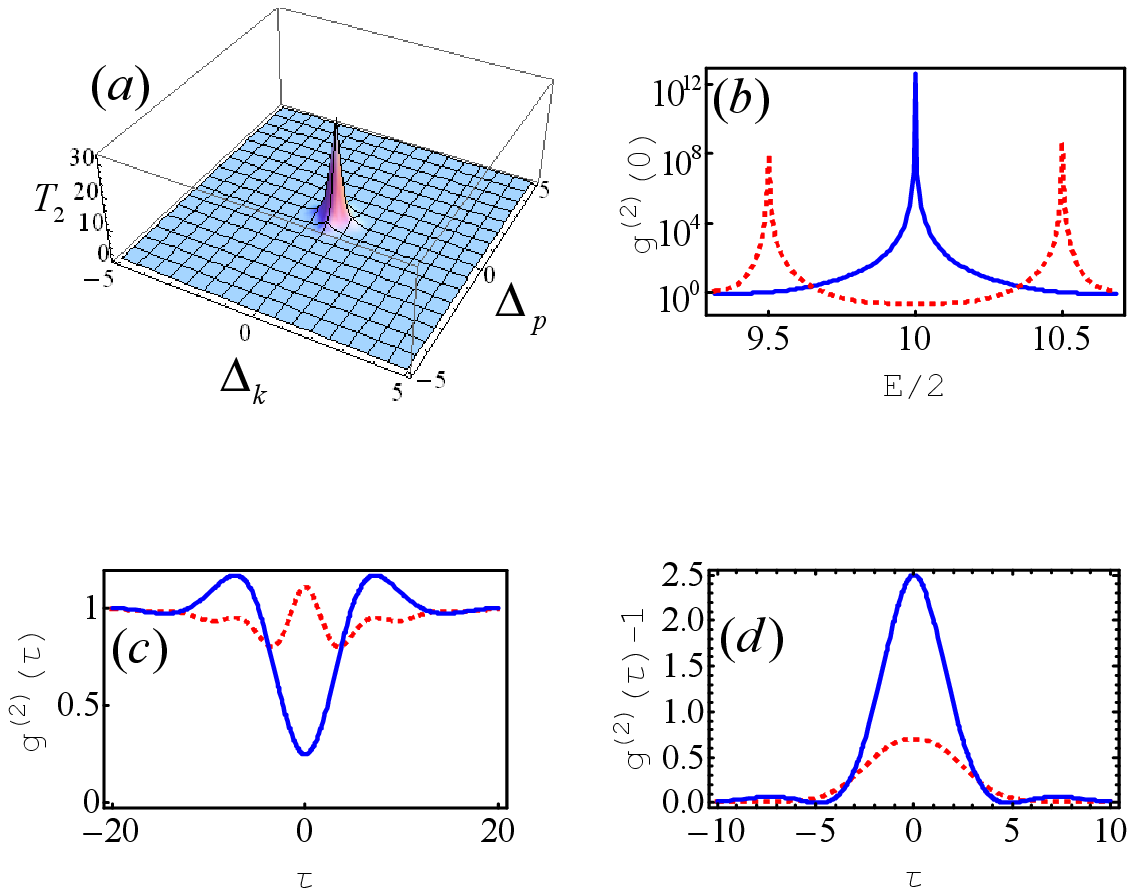}
\caption{(Color online) The two-photon background fluorescence and the
correlation functions for weak coupling regime $g=0.5$, the system
parameters are the same as that in Fig. 2: (a)-(c) represent the same
physical meaning in Fig. 2; (d) $(g^{(2)}(\protect\tau )-1)/10^{11}$ of the
reflected (solid (blue) curve) photons for $E/2=\Omega $ and $(g^{(2)}(%
\protect\tau )-1)/10^{10}$ of transmitted (dashed (red) curve) photons for $%
E/2=\protect\lambda _{1s}^{(0)}$.}
\label{fig3}
\end{figure}

In the side-coupled structure considered here, the reflected light arises
purely from the decaying amplitudes from the cavity. Thus, single photon
reflection peaks at a single photon energy of $\lambda _{1\pm }^{(0)}$,
which is the energy level for one-photon dressed state in the cavity.
Similarly, two-photon reflection peaks when the two photon energy is at $%
\lambda _{2\pm }^{(0)}$, where the cavity supports two-photon dressed
states. However, since $\lambda _{2\pm }^{(0)}\neq 2\lambda _{1\pm }^{(0)}$,
two photons each with energy $E/2=\lambda _{1\pm }^{(0)}$ is off resonance
from the two-photon dressed state. In such a case, the single excitation by
the first photon in fact prevents the second photon from entering the
cavity, resulting in the photon-blockade effect. Therefore, in contrast to
the case with direct coupling where the photon-blockade effect manifests as
a vanishing two-photon transmission, in the side-cavity case the photon
blockade effect manifests as a vanishing two-photon reflection effect. When $%
E/2=\Omega $, the single-photon reflection coefficient vanishes \cite{RS}.
The two-photon reflection is due purely to the correlation induced by the
TLS, which creates a two-photon bound state, and hence generates a large
bunching effect \cite{B}. We, therefore, for the first time, provide an
exact analytic formula for the photon correlation function for this system,
the special case of which agrees with the existing experimental data.

\textit{Weak coupling regime. --- }For the weak coupling regime with $g<%
\tilde{V}^{2}$, the two-photon background fluorescence $T_{2}$ and the
correlation function $g^{(2)}(\tau )$ are shown in Fig. \ref{fig3}. Fig. \ref%
{fig3}a shows that the four peaks in Fig. \ref{fig2}a merge into a single
peak when $E=\lambda _{2\pm }^{(0)}$.

We find that $g^{(2)}(0)$ of two reflected photons (Fig. \ref{fig3}b) has a
simple structure with one peak at $E/2=\Omega $ and always satisfies $%
g^{(2)}(0)\geq 1$, which means that the statistics of reflected photons can
not be sub-Poissonian and the photon blockade effect vanishes due to the
small energy splitting of $\lambda _{1\pm }^{(0)}$ for the weak coupling $g$%
. The bunching behaviors exhibited by $g^{(2)}(\tau )$ of two reflected
photons for $E/2=\lambda _{1\pm }^{(0)}$ (dashed (red) curve in Fig. \ref%
{fig3}c) and $\Omega $ (solid (blue) curve in Fig. \ref{fig3}d) also
illustrate the vanishing of photon blockade effect. In addition, $%
g^{(2)}(\tau )$ of anti-bunched and bunched transmitted photons for $%
E/2=\Omega $ and $\lambda _{1\pm }^{(0)}$ are shown by the solid (blue)
curve in Fig. \ref{fig3}c and the dashed (red) curve in Fig. \ref{fig3}d.

\textit{Conclusion. --- }We have\ analytically\ studied the two-photon
transport in a waveguide coupled to the cavity containing TLS, and obtained
an exact analytic formula describing the photon blockade effect. The quantum
statistics of the outgoing photons are discussed in details using the exact
two-photon wave-functions. These results agree with the observations of
recent experiment \cite{Kimble}. Our theoretical approach can also be
generalized to the many-photon transport cases.

The work is supported by National Natural Science Foundation of China a
under Grant No. 10874091 .

\end{document}